\documentclass[11pt,preprint]{aastex}

\shorttitle{UDG}
\shortauthors{Burkert}

\begin{document}

\title{The geometry and origin of ultra-diffuse ghost galaxies}

\author{A. Burkert\altaffilmark{1,2}}

\altaffiltext{1}{University Observatory Munich (USM), Scheinerstrasse 1, 81679 Munich,
Germany}
\altaffiltext{2}{Max-Planck-Fellow,
Max-Planck-Institut f\"ur extraterrestrische Physik (MPE), Giessenbachstr. 1, 85748 Garching, Germany}

\email{burkert@usm.lmu.de}

\newcommand\msun{\rm M_{\odot}}
\newcommand\lsun{\rm L_{\odot}}
\newcommand\msunyr{\rm M_{\odot}\,yr^{-1}}
\newcommand\be{\begin{equation}}
\newcommand\en{\end{equation}}
\newcommand\cm{\rm cm}
\newcommand\kms{\rm{\, km \, s^{-1}}}
\newcommand\K{\rm K}
\newcommand\etal{{\rm et al}.\ }
\newcommand\sd{\partial}

\begin{abstract}
The geometry and intrinsic ellipticity distribution of ultra diffuse galaxies
(UDG) is determined from the line-of-sight distribution of axial ratios q of a large sample
of UDGs, detected by Koda et al. (2015) in the Coma cluster. With high significance the
data rules out an oblate, disk-like geometry, characterised by major axi a=b$>$c. 
The data is however in good agreement with prolate shapes, corresponding to a=b$<$c. 
This indicates that UDGs are not thickened, rotating, axisymmetric disks, puffed up by 
violent processes. Instead they are anisotropic elongated cigar- or bar-like structures,
similar to the prolate dwarf spheroidal galaxy population of the Local Group. The 
intrinsic distribution of axial ratios of the Coma UDGs is flat
in the range of $0.4 \leq $a/c$ \leq 0.9$ with a mean value of $\langle a/c \rangle = 0.65 \pm 0.14$.
This might provide important constraints for theoretical models of their origin. 
Formation scenarios that could explain the extended prolate nature of UDGs are discussed. 
\end{abstract}


\section{Introduction}
A new class of rather peculiar but frequent galaxies has been discovered, 
called Ultra-Diffuse Galaxies (UDG; van Dokkum et al. 2015a,b; Koda et al. 2015; van der Burg, Muzzin \& Hoekstra 2016;
Roman \& Trujillo 2016). UDGs are quiescent stellar systems on the red galaxy sequence with exceptionally
low stellar surface densities of order a few M$_{\odot}$ pc$^{-2}$, two orders of magnitudes below the typical surface densities
of Milky-Way type objects. They have effective radii of 2-6 kpc, similar to giant galaxies. Their luminosities
are however of order $10^8$ L$_{\odot}$, resembling dwarf galaxies. UDGs had been seen earlier
(Impey, Bothun \& Malin 1988, Dalcanton et al. 1997). Only recently has it however become clear that they can be quite
ubiquitous, contributing significantly to the total galaxy population at least in some galaxy clusters. This was shown by
Koda et al. (2015; see also Yagi et al. 2016), who detected of order 800 UDGs in the Coma cluster with a radial distribution
within the cluster that is similar to the giant galaxies outside a core radius of 300 kpc.

The origin of these ghost galaxies represents an interesting mystery. 
Gas, entering a dark halo  will always tend to settle into a rotationally supported disk which is the 
lowest energy state for given specific angular momentum. The stars that form from this gas disk should then also 
be distributed in a disk.  Amorisco \& Loeb (2016) argue that the large radii of UDGs might result
from the infall of very high-angular momentum gas. The 
stellar surface mass densities of UDGs are however smaller than the critical gas surface density, of order 
10 M$_{\odot}$ pc$^{-2}$, required for molecular gas clouds to form and condense into stars
(McKee \& Krumholz 2010). This indicates that the surface densities of the gas disks during the star formation phase
were much higher than the observed stellar surface densities. It in turn means that the
current diffuse state results at least partly from a phase of
substantial gaseous mass loss, e.g. through ram pressure stripping or stellar feedback driven galactic winds 
(Agertz \& Kravtsov 2015; Yozin \& Bekki 2015), accompanied maybe by an expansion of the stellar disk. 
Di Cintio et al. (2016) show that multiple episodes of gas infall and blow-out could result in such
an expansion of the stellar disk and, interestingly, the formation of a cored dark matter halo (Burkert 1995).
This is in agreement with earlier numerical simulations of Ogiya \& Mori
(2011, 2014) and Ogiya et al. (2014) who find that the non-linear response of dark halos to periodic 
events of gas in- and outflows could generate cored dark matter density distributions with scaling relations
that are in agreement with observations (e.g. Burkert 2015; Kormendy \& Freeman 2016).
Violent destruction of an early, fast rotating disk would also be consistent with the 
measurements of high stellar velocity dispersions in one of the largest Coma UDGs, Dragonfly 44 (van Dokkum et al. 2016; 
see also Mart\'inez-Delgado et al. 2016). Its ratio of velocity dispersion 
to rotational velocity is 0.2 and its mass-to-light ratio is 50 -100, indicating that the stellar system is
not self-gravitating. The stars are just tracer particles within a dominant dark halo potential,
despite the fact that at the time of their formation the gas disk must
have been self-gravitating in order to form stars.
Because if its kinematics, van Dokkum et al. (2016) classified this galaxy as a dispersion dominated,
elliptical-like galaxy, rather than a disk galaxy. 
Interestingly, the individual galaxies studied by van Dokkum et al. (2016)
and Mart\'inez-Delgado et al. (2016) live in low density field environments. The harsh conditions
of a galaxy cluster might therefore not always be required for their formation. On the other hand,
Mihos et al. (2015) discovered one UDG in the Virgo cluster that shows evidence for tidal disruption.
Most UDGs however do not show such signatures and
might have been shaped by internal processes, probably a combination of high-angular momentum gas infall
(Amorisco \& Loeb 2016), combined with one or more violent episodes of gas 
loss (Di Cintio et al. 2016).

If the stellar component of UDGs formed in a dense, self-gravitating disk-like gas component,
the question arises whether information of this initial state is still hidden in the current structure of 
UDGs. Galactic disks usually are characterised by low Sersic indices n, of order unity. Indeed the surface density distribution
of UDGs is well described by n $\approx 1$. Van Dokkum et al. (2015a) and Roman \& Trujillo (2016) however argue
that the distribution of apparent axial ratios q is not flat, as expected for thin disks. UDGs could however be thick
disks, puffed-up by perturbations as discussed earlier. Typical axial ratios
are quite high with q $\approx 0.5$. They are therefore more likely spheroids than thick disks. 

Thick, axisymmetric disks resemble oblate spheroids. The observed
ellipticities could however also be explained by prolate shapes, as e.g. expected if the UDGs are bars or 
anisotropic ellipticals, affected by 
tidal fields or if the stars just trace the potential of a prolate dark matter core. 
Determining the intrinsic geometry and the true intrinsic ellipticity distribution
of UDGs could therefore add valuable information and constraints for a better understanding of their origin.

This is the purpose of this paper. Koda et al. (2015) provide a histogram of the measured axial ratios
of a large sample of UDGs in the Coma cluster. In Section 2 we deproject an updated histogram of the axial
ratio distribution of Coma UDGs, provided by Jin Koda (private communication, see also Koda et al. 2015) and derive their true
intrinsic axial ratios. We demonstrate that, with high significance, the  Coma UDG population cannot have
an oblate, disk-like geometry. Adopting prolate shapes, we find an intrinsic ellipticity distribution
that in projection is in good agreement with the observations. A discussion of this result and conclusions follow
in Section 3.

\section{Deciphering the intrinsic geometrical shape and ellipticity distribution of Coma UDGs}

Determining the distribution of true axial ratios $\beta$ or ellipticities $\epsilon = 1 - \beta$ from
the known distribution of apparent axial ratios q has a long history (e.g. Fall \& Frenk 1983; Lambas, Maddox \& Loveday 1992). 
One generally assumes concentric, spheroidal isophotes with their coordinates (x,y,z) following the relation

\begin{equation}
\frac{x^2}{a^2} + \frac{y^2}{b^2} + \frac{z^2}{c^2} = 1
\end{equation}

\noindent a,b and c are the three major axi of the spheroid. 
Here we will restrict ourselves to the two most simple possibilities:
oblate, disk-like spheroids with a = b  $\geq$ c and axial ratios $\beta$ = c/a or elongated, bar-shaped, prolate
spheroids with a = b  $<$ c and axial ratios $\beta$ = a/c. 
The following Monte-Carlo method can be used in order to determine the apparent axial-ratio distribution N(q) from random
projection of an ellipsoid with intrinsic axial ratio $\beta$. First, a projection angle $\theta$ has to be chosen, uniformly
distributed in $\sin(\theta)$ on $[0,1]$. The projected axial ratio q is then given by (Binney \& Merrifield 1998; section 4.3.3)

\begin{equation}
q = \left(\beta^2 \sin^2 \theta + \cos^2 \theta \right)^{1/2}
\end{equation}

\noindent for oblate geometries and by

\begin{equation}
q = \left( \frac{\sin^2 \theta}{\beta^2} + \cos^2 \theta \right)^{-1/2}
\end{equation}

\noindent for prolate objects.

\begin{figure}[!ht]
\epsscale{0.6}
\plotone{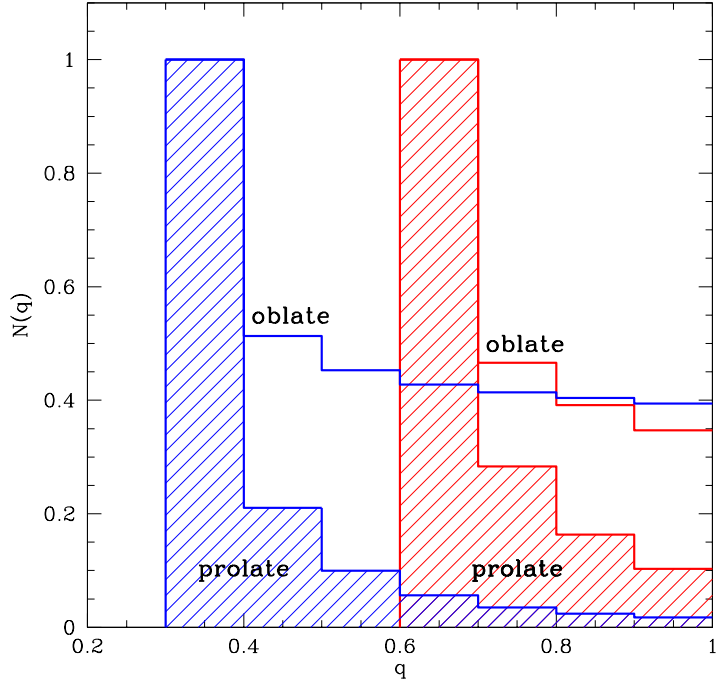}
\caption{
 \label{fig1}
Projected axial ratio distributions derived for randomly projected oblate (solid lines) 
and prolate (histograms) spheroids with intrinsic axial ratios of $\beta = 0.3$ (blue)
and $\beta = 0.6$ (red), respectively. 
The distribution of apparent axial ratios was binned in the same
way as the observed Coma UDGs.}
\end{figure}

\noindent Figure 1 shows normalized, apparent axial ratio distributions N(q) for two prolate and oblate ellipsoids with 
$\beta = 0.3$ and $0.6$. The distribution has a sharp peak at q=$\beta$. Obviously there are no galaxies
with q $< \beta$. For prolate spheroids (dashed curves)
N(q) decreases continuously towards q=1. For oblate spheroids, however, N(q) reaches a high plateau.
Oblate spheroids of all axial ratios $\beta$ therefore can contribute significantly to the observed
population of UDGs that in projection appear round (q $\approx$ 1). 

\begin{figure}[!ht]
\epsscale{1.05}
\plotone{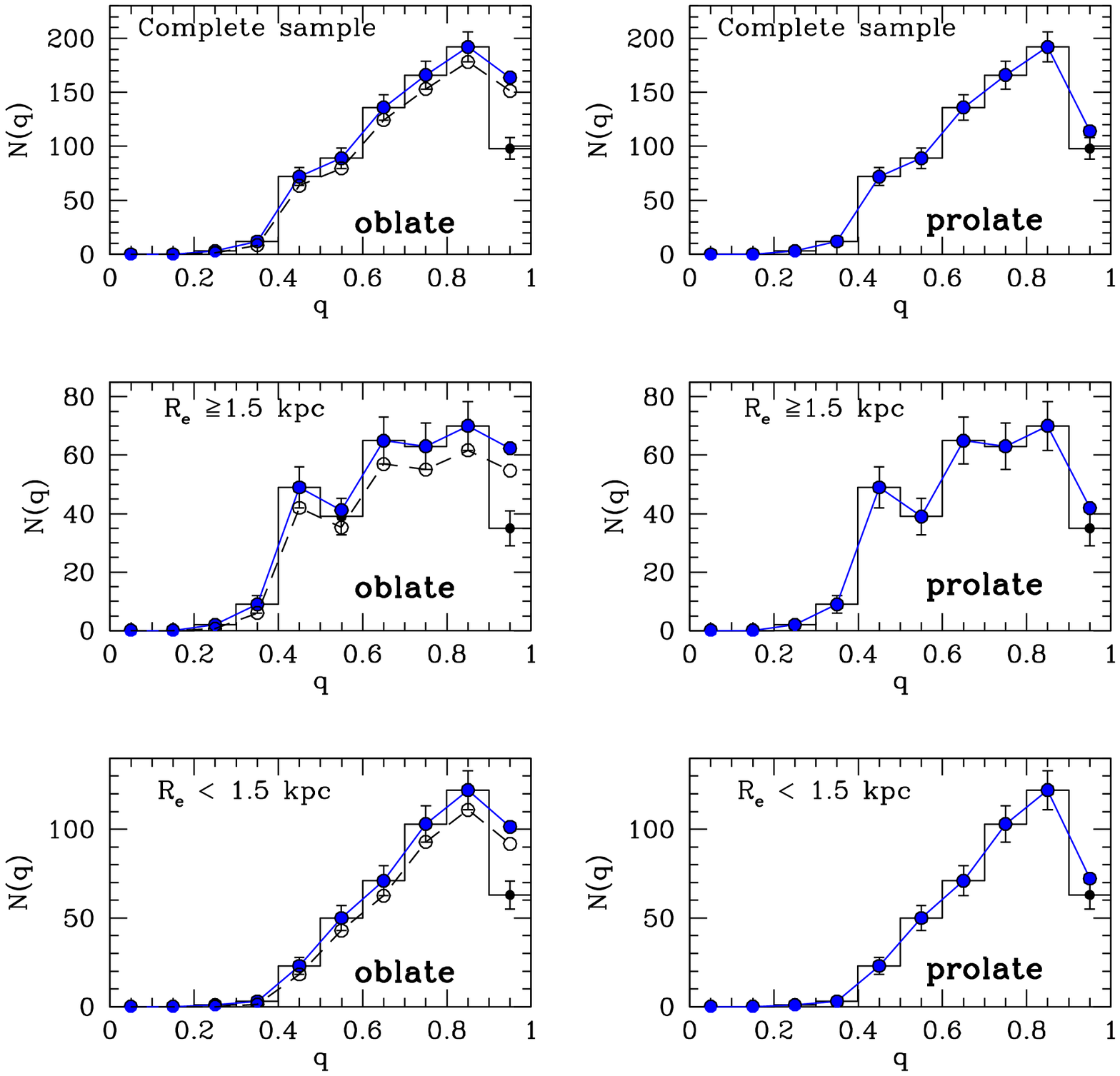}
\caption{
 \label{fig2}
The histograms in each panel show the observed distribution of apparent axial ratios q
of the full sample of Coma UDGs (upper panel), the subsample of large UDGs, characterised by
effective radii R$_e \geq$ 1.5 kpc (middle panel)
and the subsample of compact UDGs  with R$_e <$ 1.5 kpc (lower panel).
Filled black circles with statistic error bars show
the observed number of galaxies. Blue filled circles correspond to the predicted apparent q distribution of 
oblate galaxies (left panels) and prolate galaxies (right panels) with an intrinsic distribution 
as determined from the deprojection method. The intrinsic axial ratio distributions
N($\beta$) for the full sample, the extended and the compact subsamples is shown in Figure 4.
The prolate deprojection method can explain the observed
q distributions well, even for the highest q bin (q = 0.95). The situation is different for the oblate sample. 
It leads to too many round galaxies in the highest axial ratio bin. Even reducing the number of oblate
UDGs in each bin by one $\sigma$ (dashed curves and open black circles) cannot explain the highest q bin.}
\end{figure}

The dominant peak at q=$\beta$ allows us to apply the following method in order to determine the intrinsic axial ratios and
the geometry of the Coma UDGs. The histograms in Figure 2 show the observed distribution 
N$_0$(q) of apparent axial ratios for the full sample (upper panels) of Coma UDGs, observed by Koda et al. (2015).
The middle and lower panels show the
subsample of extended Coma UDGs with effective radii R$_e \geq 1.5$ kpc and compact UDGs with  R$_e < 1.5$ kpc, respectively.
We begin with the lowest q bin where N$_0$(q) $>$ 0. In our case this corresponds to the q=0.25 bin. We now
determine the randomly projected axial ratio distribution N(q$|$0.25) of galaxies with intrinsic axial ratio $\beta = 0.25$. 
This projected distribution is then binned in q similar to the observations and normalized such that for q=0.25 the number
N(0.25$|$0.25) is equal to the observed number of UDGs in this bin, i.e. N(0.25$|$0.25) = N$_0$(0.25). 
We now can determine the distribution of not yet assigned UDGs by subtracting N(q$|$0.25) from N$_0$(q).
This leads to a new distribution N$_1$(q) = N$_0$(q) - N(q$|$0.25). N$_1$(0.25) = 0, but N$_1$(0.35) $>$ 0. So
we can proceed to the q=0.35 bin and repeat the same procedure.
First we determine the projected distribution N(q$|$0.35) of galaxies with $\beta$ = 0.35 which is then
normalized such that N(0.35$|$0.35) = N$_1$(0.35). 
We subtract N(q$|$0.35) from N$_1$(q) which leads to the distribution N$_2$(q) which allows us to determine the number of UDGs
in the intrinsic axial ratio bin $\beta = 0.45$. We continue this iteration till we reach the final bin, corresponding
to q=0.95.  As shown by the filled blue points in Figure 2, this procedure allows us to fit the observations 
perfectly up to this last bin, independent of whether we adopt an oblate or prolate 
geometry. Never does the subtraction produce a negative number for N$_i$(q) within the statistical error bars, 
which would be unphysical. 
This results from the fact that the observed distribution continuously 
increases with increasing q. For the sample of large galaxies with R$_e \geq 1.5$ kpc, there is a drop in the 
observed number at q=0.55. This does not lead to a problem for prolate geometries. However, adopting oblate shapes, 
random projection of the UDGs with lower q leads to 41 UDGs in this q bin while only 39 are observed. 
As shown by the corresponding filled blue point in the q=0.55 bin, this
minor difference of 2 is still within the statistical error bar. 

The situation is however very different for the last, highest axial-ratio bin, where
a strong drop in number is observed. The blue point in the q=0.95 bin of the upper left panel of Figure 2
shows the predicted number of galaxies in this bin {\it just from projection effects of galaxies in bins with smaller q values}, 
assuming an oblate geometry.
Note that we did not even add any additional galaxies with intrinsic axial ratios $\beta$=0.95.
The predicted number of projected round UDGs is 164, far too large, compared to 
the 98 observed round galaxies. Even if we would reduce the number of observed galaxies in each bin with q $<$ 0.95
by one sigma (open black circles and dashed line) would we end up with an uncomfortably large number of 151 round galaxies,
resulting just from projection effects. The same is true for the extended and compact subsamples (middle and
lower left panels of Figure 2).
The situation is however very different for prolate geometries. As projections parallel to the long
axis are very unlikely (see Figure 1) for this geometry, the number of apparently round galaxies is much smaller. 
For the full sample (upper right panel of Figure 2) we now find
just 114 round UDGs, resulting from projection effects which is almost within one sigma of the observed number.

\begin{figure}[!ht]
\epsscale{0.9}
\plotone{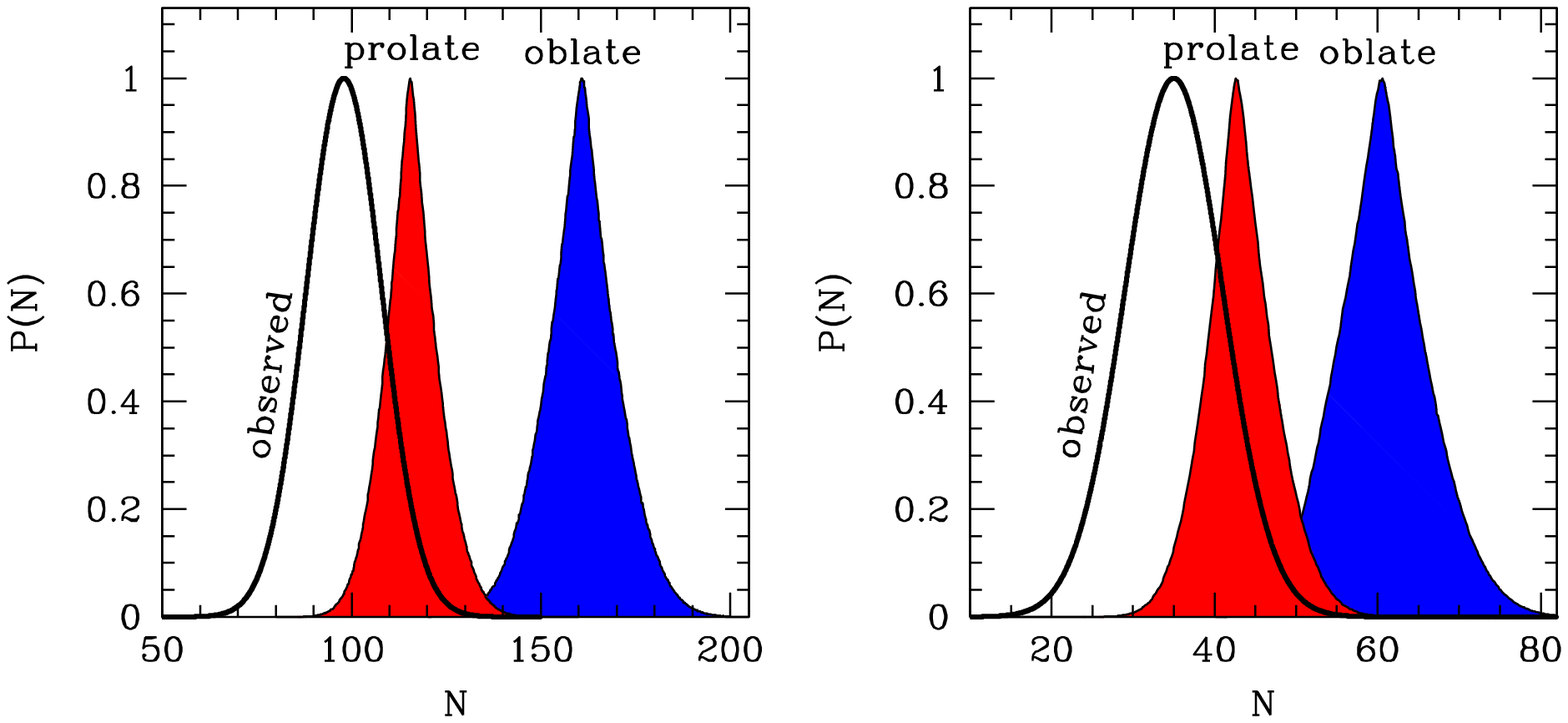}
\caption{
 \label{fig3}
The observationally inferred number distribution of round UDGs, that is UDGs
in the highest q=0.95  bin, is shown by the solid black line. 
The red and blue distributions show the predicted number of round galaxies 
that one would expect just from random projection of all the galaxies in the other q bins,
adopting an oblate (blue) or prolate (red) geometry. The left panel shows the full sample.
The distribution of extended UDGs is shown in the right panel.}
\end{figure}

In order to quantify the significance of this result another Monte-Carlo simulation was applied.
For each bin with q $<$ 0.95 we choose a number of galaxies which follows a normal distribution of
projected axial ratios that peaks at the observed number with a width given by the statistical $\sqrt{N}$ error.
As before we then determine the number of galaxies with projected axial ratios in the q = 0.95 bin. Repeating this procedure many times
leads to a distribution of the predicted number of projected round (q=0.95) galaxies that is shown
in Figure 3. The solid line corresponds to the observations (N=98 $\pm$ 10), adopting a normal distribution with statistical errors.
The red shaded area shows the distribution, expected for prolate geometries which overlaps with the observations
within one $\sigma$.  Adopting oblate shapes however produces the blue shaded distribution which is many $\sigma$ off and can
therefore clearly be ruled out.  In summary, the apparent ellipticity distribution of Coma UDGs strongly favors a prolate,
rather than an oblate geometry. 

The red shaded regions in Figure 4 show the deprojected, intrinsic 
axial ratio distributions of the full sample of UDGs (upper panels) and of the extended 
and compact UDGs alone (middle and lower panels).
Independent of their size and the assumption of a prolate or oblate geometry
most galaxies have axial ratios in the range of $0.4 \leq a/c \leq 0.9$ with a flat distribution.
For the complete sample, mean values are $\langle a/c \rangle = 0.65 \pm 0.14$. Within the statistical uncertainties,
the extended and compact subsamples have the same mean axial ratios, with $\langle a/c \rangle = 0.68 \pm 0.15$ for 
the extended UDGs and $\langle a/c \rangle = 0.61 \pm 0.15$ for the compact galaxies.

Up to now we adopted the 10 projected axial ratio bins of the updated sample, provided by Jin Koda (private communication).
Individual axial ratios for each galaxy were not available for this sample. In order to test the dependence of the result
on binning, the table of observed axial ratios, published in Koda et al. (2015) was adopted. For $N_q=10$ bins the previous
results are recovered. The situation does not change if we vary the number of bins in the range of $N_q=6$ to $N_q=18$. 
In all cases, prolate geometries provide a good fit to the data while oblate geometries wastly overpredict
the number of round galaxies. For smaller values of $N_q$ the strong decline in the number of round objects
is smeared out. For larger bin numbers statistical fluctuations become too large
in order for this model to converge and produce a predicted number of round galaxies from random projections
of galaxies with larger axial ratios.

\begin{figure}[!ht]
\epsscale{0.9}
\plotone{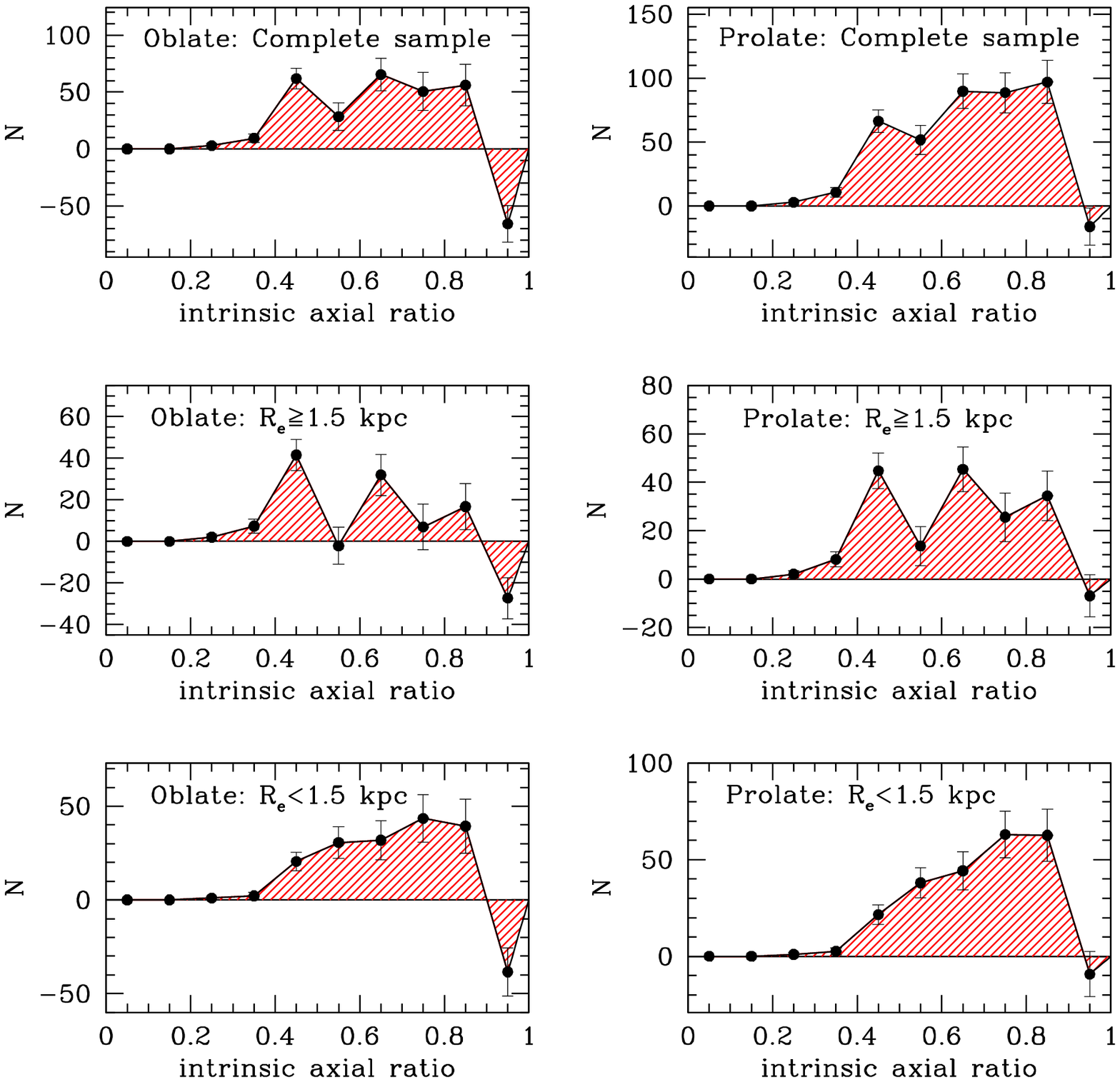}
\caption{
 \label{fig4}
The intrinsic axial ratio distribution of UDGs is shown for oblate (left panels)
and prolate (right panels) geometries. The upper panels show the full sample. The middle and 
lower panels depict the population of extended UDGs with effective radii R$_e \geq $ 1.5 kpc and
of compact UDGs with R$_e < $ 1.5 kpc, respectively.}
\end{figure}

\section{Discussion and Conclusions}

The deprojection of observed axial ratios of Coma UDGs leads to the conclusion that these galaxies are on average
prolate, rather than oblate. Focussing on Figure 4 and adopting an oblate geometry (upper left panel), 
N is strongly negative for the largest axial ratio bin. The problem is much smaller for prolate geometries. One could of course
argue that this large observed deficit of round galaxies, adopting oblate intrinsic shapes, is a result of some observational bias, 
such that oblate spheroids, seen face on, are more difficult to detect.
In this case, at least 66 round UDGs, out of a total number of 768 UDGs, would be missing which appears unlikely.
One could also argue that UDGs have a mixture of shapes, some being oblate, others prolate.
Note however that even for prolate shapes projection effects lead to a somewhat larger number of round galaxies,
compared with the observations. This $\sim 1 \sigma$ discrepancy exists independent of binning and might indeed
indicate some observational bias against round ellipticals.
Assuming that some of these galaxies are actually oblate would however make this problem worse. 
Minimizing the difference between observed round galaxies and theoretically expected round galaxies therefore requires 
that all UDGs are prolate. 

Interestingly, Coma UDGs and Virgo cluster dwarf ellipticals have a very similar distribution of shapes.
Chen et al. (2010, see also Lisker et al. 2009) investigated the structural properties of 100 ACS Virgo
Cluster Survey galaxies. The mean axial ratios of their faint galaxy sample is $\langle$ q $\rangle = 0.73 \pm 0.18$.
This value is precisely the same as the projected mean axial ratio of the Coma 
UDGs with $\langle$ q $\rangle = 0.72 \pm 0.16$. In addition, Chen et al. (2010) find no Virgo dwarfs,
flatter than 0.35. And, like the Coma UDGs, the number of dwarfs increases with increasing
axial ratios, with a depression in the highest axial ratio bin, that is for round objects. Chen et al. (2010) note
that this signature is not consistent with random projection of flat, disk-like geometries.
Similar processes might therefore have shaped the Coma UDGs and the Virgo dwarf galaxy population.

That UDGs are prolate, rather than oblate should provide important information about the processes
that lead to their characteristic diffuse state. First of all, it indicates that they are dispersion dominated bars
or cylinders, rather than rotation supported, thick disks, in agreement with the small $v_{rot}/\sigma$ values found by 
van Dokkum et al. (2016). If the stellar component is
in virial equilibrium within a dominant dark halo potential this could be achieved either by assuming that the dark halo
distribution itself is prolate and/or that the stellar system's velocity dispersion is anisotropic with larger
dispersions parallel to the long axis, probably as a result of the processes that generated their elongated state.
In this case, the population of projected round UDGs should have systematically larger observed velocity dispersions 
than flattened UDGs of the same stellar mass, as the line-of-sight of round UDGs would be parallel to the long axis. 
The situation would be the opposite
for oblate galaxies where velocity dispersions are usually larger in the equatorial plane than perpendicular to it.

It is interesting that prolate shapes would link the UDGs to their diffuse, low-mass relatives:
Local Group dwarf spheroidal galaxies (dSphs).
Hayashi \& Chiba (2015), for example, investigated the kinematical data of 12 dSphs and argue that these objects probably
life in elongated, bar-like dark matter halos with their shapes reflecting the elongation of their confining dark halos.
They also note that, if true, this elongation is inconsistent with $\Lambda$CDM models. This is interesting as it might indicate
that internal processes that lead to the diffuse state of these galaxies also reshaped their inner dark halo components.

The flat distribution of prolate axial ratios of UDGs in the range of 0.4 to 0.9, independent of their effective radii, 
provides important constraints for any theoretical model of their formation. 
Up to now, we do not know whether existing models can reproduce this result. For example, Ceverino et al. (2015) and
Tomassetti et al. (2016) find in their high-resolution numerical simulations that prolate shapes of stellar systems 
and dark matter halos are generic for lower-mass galaxies with stellar masses of order M$_* \leq 10^9$ M$_{\odot}$. 
These authors focussed on early galaxies in the high cosmic redshift regime of z $\approx 2-4$. 
Whether this applies also to present-day galaxies and to the peculiar and extreme
physical conditions that produce UDGs is however not clear. In addition, it has not been shown that the distribution is flat and
in the observed range of axial ratios.

If the stars formed in a fast rotating disk that was lateron puffed up by substantial mass loss one might expect
that the system keeps a preferentially oblate shape. On the other hand, violently relaxing particle systems can 
experience radial orbit instability that generates a bar with an anisotropic velocity dispersion (Merritt 1987).
Another process could be an encounter with a massive perturber or tidal effects caused by the Coma cluster potential.
 An interested suite of numerical simulations of tidally stirred disky satellite galaxies has been presented by
Martinez-Delgado et al. (2013, see also Kazantsidis et al. 2017). 
They focussed on the population of  dwarf spheroids orbiting Milky-Way-sized hosts and
demonstrated that especially for cored dark matter halos (Burkert 1995) tidal effects could transform
oblate, disky dwarfs into prolate spheroids. The same mechanism might actually be active for UDGs in cluster potentials
although it is not clear yet whether it would lead to the observed ellipticity distribution with
$\langle a/c \rangle = 0.65 \pm 0.14$. This scenario is also
promising, as it might explain the peculiar orientiation of the UDG's long axis in Coma 
which is not random. Yagi et al. (2016) find that the UDG's major axis is preferentially radially aligned towards the cluster center.
Note however, that UDGs have now also been found in low-density environments. Martinez-Delgado et al. (2016) report
the discovery of an ultra-diffuse, quenched galaxy, DGSAT 1, in the outskirts of the Pisces-Perseus supercluster. They
argue that DGSAT 1 might be a "backsplash" galaxy that passed through the center of the cluster Zw 0107+3212 (Gill et al. 2015)
with high velocities and is now in the outskirts at distances of 1-2 virial radii. If field UDGs did not experience
tidal effects they might actually be oblate, rather than prolate. In order to test this conjecture one would however need
a sample of order a few 100 objects in order to determine their geometrical shape using the method, presented in this paper.

Maybe, the stars in UDGs never formed in a disk configuration. Could star formation have occured in 
individual dense clouds, orbiting within the inner region of a cored, prolate dark halo? 
Like normal galaxies, UDGs have an extended globular cluster 
population (Beasley et al. 2016; Peng \& Lim 2016; Beasley \& Trujillo 2016; van Dokkum et al. 2016), similar
to the halo globular clusters of the Milky Way.  Did the stars and globulars form
ex-situ in substructures that were lateron accreted by a dark halo that did not manage to form in-situ stars?
In this case, UDGs would be failed galaxies with an accreted stellar halo component, similar to the Milky Way halo,
but missing a disk component. Even more extreme would be the assumption that the more than 800 Coma UDGs 
are actually galaxies that are currently in the process of being tidally disrupted (Hozumi \& Burkert 2015;
Ploeckinger et al. 2015). A prominent example of such a process
is the S-shaped dwarf galaxy, observed in the Hydra I cluster (Koch et al. 2012).

In order to distinguish between these scenarios more detailed observations, including an investigation of the kinematics and
rotation of UDGs and the metallicity and age distribution of their stellar populations is required. 

\acknowledgements

This work was supported by the cluster of excellence "Origin and Structure of the Universe". 
I thank Jean Brodie, Avishai Dekel, Duncan Forbes, Aaron Romanowsky and Pieter van Dokkum for inspiring discussions and
the astronomy department at the University of California, Santa Cruz, for their hospitality and support during the
preparation of this paper. Thanks also to Jin Koda for providing an updated histogram of Coma UDG axial ratios.
I also thank the referee for carefully reading the manuscript and for excellent comments.

\end{document}